# Back to the Moon: The Scientific Rationale for Resuming Lunar Surface Exploration

I. A. Crawford[a,b,*], M. Anand[c,d], C. S. Cockell[e], H. Falcke[f,g,h], D. A. Green[i], R. Jaumann[j], M. A. Wieczorek[k]

[a] Department of Earth and Planetary Sciences, Birkbeck College, Malet Street, London, WC1E 7HX, UK.

[b] Centre for Planetary Sciences at UCL/Birkbeck, UK.

[c] Planetary and Space Sciences, Department of Physical Sciences, The Open University, Milton Keynes, MK7 6AA, UK.

[d] Department of Mineralogy, The Natural History Museum, London, SW7 5BD, UK.

[e] School of Physics and Astronomy, University of Edinburgh, UK.

[f] Department of Astrophysics, Radboud University, Nijmegen, The Netherlands.

[g] Netherlands Institute for Radio Astronomy (ASTRON), Postbus 2, 7990 AA Dwingeloo, The Netherlands.

[h] Max-Planck-Institut für Radioastronomie, Auf dem Hügel 69, 53121 Bonn, Germany.

[i] Centre of Human and Aerospace Physiological Sciences, King's College London, London, SE1 1UL, UK.

[j] DLR, Institute of Planetary Research, Berlin, Germany.

[k] Institut de Physique du Globe de Paris, Univ Paris Diderot, France.

[*] Corresponding author: Tel : +44 203 073 8026

   Email address: i.crawford@bbk.ac.uk

**Abstract**

The lunar geological record has much to tell us about the earliest history of the Solar System, the origin and evolution of the Earth-Moon system, the geological evolution of rocky planets, and the near-Earth cosmic environment throughout Solar System history. In addition, the lunar surface offers outstanding opportunities for research in astronomy, astrobiology, fundamental physics, life sciences and human physiology and medicine. This paper provides an interdisciplinary review of outstanding lunar science objectives in all of these different areas. It is concluded that addressing them satisfactorily will require an end to the 40-year hiatus of lunar surface exploration, and the placing of new scientific instruments on, and the return of additional samples from, the surface of the Moon. Some of these objectives can be achieved robotically (e.g. through targeted sample return, the deployment of geophysical networks, and the placing of antennas on the lunar surface to form radio telescopes). However, in the longer term, most of these scientific objectives would benefit significantly from renewed human operations on the lunar surface. For these reasons it is highly desirable that current plans for renewed robotic surface exploration of the Moon are developed in the context of a future human lunar exploration programme, such as that proposed by the recently formulated Global Exploration Roadmap.

# 1. Introduction

Over the last decade there has been a renaissance in lunar exploration conducted from orbit about the Moon, with the following countries all sending remote-sensing spacecraft to lunar orbit during this period: European Space Agency: SMART-1 (2004); Japan: Kaguya (2007); China: Chang'e-1 (2007), Chang'e-2 (2010); India: Chandrayaan-1 (2008); and the United States: Lunar Reconnaissance Orbiter (LRO; 2009), GRAIL (2012). However, none of these spacecraft were designed to land on the Moon's surface in a controlled manner, although the US Lunar CRater Observation and Sensing Satellite (LCROSS; co-launched with LRO) and the Chandrayaan-1 Moon Impact Probe (MIP) did deliberately impact the lunar surface in an effort to detect polar volatiles. Indeed, it is sobering to realise that no spacecraft of any kind has successfully landed on the Moon in a controlled manner since the Russian robotic sample return mission Luna 24 in August 1976, and that no human being has set foot on the Moon since Apollo 17 in December 1972. In this paper we argue that this long hiatus in lunar surface exploration has been to the detriment of lunar and planetary science, and indeed of other sciences also, and that the time has come to resume the robotic and human exploration of the surface of the Moon.

The strong scientific case for renewed lunar exploration was already recognized in a comprehensive study conducted by the European Space Agency in 1992 on "Europe's Priorities for the Scientific Exploration and Utilization of the Moon" (ESA 1992). Many of the conclusions of this earlier study remain valid today, albeit with some need of updating in light of more recent discoveries, and we will follow its basic framework in what follows. In particular, the ESA study found that lunar science objectives can logically be divided into three categories: (i) Science of the Moon (i.e.

studies of the Moon itself); (ii) Science on the Moon (i.e. studies using the lunar surface as a platform for scientific investigations not directly related to the Moon itself); and (iii) Science from the Moon (i.e. studies utilising the lunar surface as a platform for astronomical observations).

## 2. Science of the Moon

From a planetary science perspective, the primary importance of the Moon arises from the fact that, owing to a lack of recent geological activity and active erosional processes, it has an extremely ancient surface, mostly older than 3 billion years with some areas extending almost all the way back to the origin of the Moon 4.5 billion years ago (e.g. Hiesinger and Head 2006; Jaumann et al., 2012). Its relatively accessible near-surface environment therefore preserves a record of the early geological evolution of a terrestrial planet, which more complex planets such as Earth, Venus and Mars, have long lost, and of the Earth-Moon system in particular. Moreover, the lunar surface provides a platform for geophysical instruments (e.g. seismometers and heat-flow probes) to probe the structure and composition of the deep interior, which is also required if we are to use the Moon as a model for terrestrial planet evolution. Last, but not least, the Moon's outer layers also preserve a record of the environment in the inner Solar System (e.g. meteorite flux, interplanetary dust density, solar wind flux and composition, galactic cosmic ray flux) throughout Solar System history, much of which is relevant to understanding the past habitability of our own planet (e.g. Crawford, 2006; Cockell, 2010).

Gaining access to this information lies primarily in the domain of the geosciences (i.e. geology, geophysics, and geochemistry), and will require a return to the lunar surface with robotic and/or human missions.

The most recent study of 'Science of the Moon' objectives is the US National Research Council's Report on the Scientific Context for Exploration of the Moon (NRC 2007; hereinafter the 'SCEM Report'). This study identified, and prioritized, eight top-level scientific 'concepts' (each of which can be broken down into multiple individual science goals), and identified the capabilities that would be required of space missions designed to address them. We note that several of these objectives (i.e. those which could be addressed by a South Pole-Aitken Basin sample return mission and a lunar geophysics network) were also strongly endorsed by the recent US Planetary Science Decadal Survey (NRC, 2011). As a basis for discussion, we summarise these here (as given in Table 4.1 of NRC 2007), and in each case highlight how surface (human and/or robotic) exploration will aid in meeting these scientific objectives.

## 2.1 The Bombardment History of the Inner Solar System

The vast majority of lunar terrains have never been directly sampled, and their ages are based on the observed density of impact craters calibrated against the ages of Apollo and Luna samples (e.g. Neukum et al., 2001; Stöffler et al., 2006). However, the current calibration of the cratering rate, used to convert crater densities to absolute ages, is neither as complete nor as reliable as it is often made out to be. For example, there are no calibration points that are older than about 3.85 Ga, and crater ages younger than about 3 Ga are also uncertain (e.g. Hiesinger et al., 2012). Furthermore, longitudinal variations in the cratering rate on the surface of the Moon are predicted by orbital dynamics models, but these are difficult to confirm with existing data (Le Feuvre and Wieczorek, 2011). Improving the calibration of the cratering rate would be of great value for planetary science for the following three reasons: (i) It would provide better estimates for the ages of unsampled regions of the lunar surface; (ii) It would provide us with a more reliable estimate of the impact history of the inner Solar System, especially that of our own planet; and (iii) The lunar impact rate is used, with various other assumptions, to date the surfaces of other planets for which samples have not been obtained (including key events and stratigraphic boundaries on Mars), and to the extent that the lunar rate remains unreliable so do the age estimates of surfaces on the other terrestrial planets.

Moreover, there is still uncertainty over whether the lunar cratering rate has declined monotonically since the formation of the Moon, or whether there was a bombardment 'cataclysm' between about 3.8 and 4.0 billion years ago characterised by an unusually high rate of impacts (Hartmann et al., 2000; Stöffler et al., 2006). Clarifying this issue is especially important from an astrobiology perspective because it defines the impact regime under which life on Earth became established (e.g. Maher and Stevenson, 1988; Sleep et al., 1989; Ryder, 2003; for an illustration of this unresolved issue see Fig. 2.3 of NRC 2007).

In principle, obtaining an improved cratering chronology is straightforward: it 'merely' requires the sampling, and radiometric dating, of surfaces having a wide range of crater densities, supplemented where possible by dating of impact melt deposits from individual craters and basins (Stöffler, et al., 2006). However, in practice this is likely to require the implementation of multiple missions to many different sites (see the detailed discussion by Stöffler, et al., 2006). These might be robotic missions, ideally involving rover-facilitated mobility and either *in situ* radiometric dating or, preferably, a sample return capability (e.g. Shearer and Borg, 2006). The possibilities of robotic *in situ* radiometric dating techniques have been investigated by a number of workers (e.g. Talboys et al., 2009; Trieloff et al., 2010), and such *in situ* measurements may be able to address some key questions in lunar science (e.g. verifying the approximate ages of the youngest lava flows estimated by crater counting). However, the practicality of performing *in situ* radiometric dating with sufficient accuracy to obtain definitive results for chronology has been questioned (e.g. Taylor et al., 2006). Additional robotic sample return missions to key localities, such as the South Pole-Aitken basin (as envisaged by the MoonRise proposal; Jolliff et al., 2010) would undoubtedly improve our knowledge in key respects (Cohen and Coker, 2010). However it is likely to prove impractical to target dedicated robotic sample return missions to sufficient locations to fully characterise lunar chronology. Given the enhanced mobility, sample collection efficiency and sample return capacity actually demonstrated by the Apollo missions (e.g. Vaniman et al., 1991; Crawford, 2012), it seems clear that obtaining anything approaching a complete lunar impact chronology will, at the very least, be greatly facilitated by new human missions to the lunar surface.

**2.2 The structure and composition of the lunar interior**

As noted in the SCEM Report, the structure of the lunar interior provides fundamental information on the evolution of differentiated planetary bodies. The Moon is especially important in this respect because, lacking plate tectonics, its crust and mantle have remained almost completely isolated from each other for more than 4 billion years. Given the limited volcanic activity on the Moon since its formation, and given the sluggish convection in the mantle of this single-plate planet, the interior of the Moon should retain a record of early planetary differentiation processes that more evolved planetary bodies have since lost.

Despite data from a rudimentary geophysical network that was set up during the Apollo missions (ALSEP; Apollo Lunar Surface Experiment Package), keys aspects of the Moon's interior structure, composition, and evolution are left unanswered (for a summary, see Wieczorek et al. 2006). The thickness of the crust, which is critical for estimating the bulk composition of the Moon, is debated, with recent estimates being almost half of those made during the Apollo era. The existence of a seismic discontinuity in the mantle 500 km below the surface, which could either be a reflection of primordial differentiation or later magmatic processes, is also debated. Four heat flow measurements (at two sites each) were made during Apollo (Langseth et al. 1976) in order constrain the Moon's bulk composition and thermal evolution, but in retrospect, these were found to be made near the edge of an atypical province enriched in heat-producing elements (Jolliff et al., 2000). It is not clear if these unique measurements are representative of either the ancient feldspathic highlands terrane, or of the more volcanically active Procellarum KREEP terrane. Though most data indicate that the Moon has a small molten core (e.g. Weber et al., 2011), neither its size nor composition are well constrained, and the existence of a solid inner core has remained elusive.

Resolving these questions will require making further geophysical measurements. While a few such measurements can be made from orbit (such as the measurement of the Moon's gravity and magnetic fields; Purucker and Nicholas 2010, Zuber et al., 2011), most require geophysical instruments to be placed on, or below, the lunar surface. Key instruments in this respect are seismometers, to probe the structure of the deep interior (e.g. Lognonné, 2005; Yamada et al., 2011), heat-flow probes to measure the heat loss from the lunar interior and its spatial variations (Langseth, et al., 1976), and magnetometers to measure the thermal conductivity profile. Such instruments can in principle be placed and operated on the lunar surface robotically without direct human intervention, such as envisaged for the proposed Farside Explorer (Mimoun et al., 2012) and Lunette (Neal et al., 2011) missions, the International Lunar Network (Cohen et al., 2010; also http://iln.arc.nasa.gov/), and the LunarNet penetrator concept (Smith et al., 2012). Nevertheless, as discussed below, human deployment of such instruments would be extremely beneficial.

The thermal perturbation caused by an autonomous lander, which would bias the heat flow measurements, could be minimized by placing the heat flow experiment far from the landing site (Kiefer, 2012). The emplacement of thermal probes several meters below the surface would be greatly facilitated by a human operated drill, as was aptly demonstrated during the deployment of the heat flow package by the Apollo astronauts. Moreover, thanks to the human operated portable magnetometers used during the Apollo 14 and 16 missions, it is known that the surface magnetic fields can vary dramatically, and even change direction, over kilometer scale distances (e.g., Fuller and Cisowski 1987); human-led magnetic surveys over large distances would be invaluable in deciphering the origins of these enigmatic fields. Strong magnetic fields can help in deflecting harmful solar and cosmic rays, but it is currently difficult to quantify this potential benefit to human exploration activities in the absence of detailed surface magnetic field map (e.g., Halekas et al., 2010).

A human return to the lunar surface would therefore benefit geophysical investigations by the additional payload capacity, operational flexibility, and long traverses that are inherent in human missions (e.g. Spudis, 2001; Garvin, 2004; Crawford, 2004).

**2.3 The diversity of lunar crustal rocks**

The SCEM Report highlighted the fact that key planetary processes are manifested in the diversity of lunar crustal rocks. Quantifying and understanding this diversity will require detailed chemical and mineralogical analysis of rocks and soils from as yet unsampled regions of the lunar surface. In particular, no samples have yet been returned from the polar regions or the far-side, thus greatly limiting our knowledge of lunar geological processes. Although, statistically, many of the 160 or so known lunar meteorites (http://meteorites.wustl.edu/lunar/moon_meteorites_list_alpha.htm) must originate from these areas, and provide an excellent statistical sampling of the lunar crust (e.g. Korotev, 2005), the value of these materials is limited by a lack of knowledge of their source regions and thus geological context.

The diversity of lunar crustal materials has been demonstrated most recently by new results from orbital remote sensing instruments on Kaguya, Chandrayaan-1 and LRO. These include outcrops of pure anorthosite which may represent pristine magma ocean flotation cumulates (Ohtake et al., 2009); olivine rich outcrops which may sample mantle material (Yamamoto et al., 2010); and spinel-rich (Sunshine et al., 2010) and silica-rich (Greenhagen et al. 2010; Glotch et al. 2010; Jolliff et al., 2011) lithologies not represented in the existing sample collection. It is important to confirm the interpretation of the remote-sensing data as no 'ground-truth' has yet been obtained for any of these localities. It is also important to obtain measurements of minor and trace elements in these materials which cannot be detected by orbital remote-sensing instruments, but which are essential to discriminate between different suggested origins and formation mechanisms.

Sample return missions to currently unsampled regions would be the preferred means of furthering our knowledge of lunar geological diversity. Although an alternative would be to make robotic *in situ* geochemical measurements, as proposed for both the Farside Explorer (Mimoun et al., 2012) and LunarNet (Smith et al., 2012) mission concepts, a full understanding of geological processes requires the ability to measure the isotopic composition and concentrations of minor and trace elements at levels that may be beyond the capabilities of *in situ* robotic instruments. In addition, mobility is crucially important if the local geological diversity in the vicinity of a landing site is to be properly characterized. As was the case for determining surface ages (Section 2.1), and for essentially the same reasons, characterizing lunar geological diversity would greatly benefit from a renewed human presence on the lunar surface provided that such missions were targeted at sites carefully chosen with this objective in mind.

**2.4 Volatiles at the lunar poles**

As noted in the SCEM Report, the lunar poles potentially bear witness to the flux of volatiles present in the inner Solar System throughout much of Solar System history.

In 1998 the *Lunar Prospector* neutron spectrometer found evidence of enhanced concentrations of hydrogen at the lunar poles (Feldman et al., 1998), which was widely interpreted as indicating the presence of water ice in the floors of permanently shadowed polar craters. Radar observations have also been used to infer the presence of ice in such regions (e.g. Spudis et al., 2010; but see Fa et al. 2011 for an alternative interpretation). The presence of water ice in permanently shadowed craters was supported by the LCROSS impact experiment, which found a water ice concentration of 5.6 ± 2.9 % by weight in the target regolith at the Cabeus crater (Colaprete et al., 2010). It seems likely that this water is ultimately derived from the impacts of comets (and/or water-rich asteroids) with the lunar surface, although solar wind implantation and endogenic sources might also contribute (for a brief review see Anand, 2010). However, the inferred quantity of water is sensitive to the calibration of the spectrometers on the LCROSS probe and a number of other assumptions (Colaprete et al. 2010). Ideally, therefore, this result needs to be confirmed by *in situ* measurements.

In addition to ice in permanently shadowed craters, infra-red remote-sensing observations have found evidence for hydrated minerals, and/or adsorbed water or hydroxyl molecules, over large areas of the high latitude (but not permanently shadowed) lunar surface (Pieters et al. 2009; Sunshine et al., 2009; Clark, 2009). It is hypothesised this $OH/H_2O$, which cannot exist as ice, is produced by the reduction of iron oxides in the regolith by solar wind-implanted hydrogen, with $OH/H_2O$ being retained in the relatively cold high-latitude regolith. It is possible that, over time, this high-latitude $OH/H_2O$ may migrate to polar cold traps and contribute to ice deposits there (Crider and Vondrak 2002), and this scenario is supported by observations of water molecules in the lunar exosphere made by the Moon Impact Probe (MIP) released by Chandrayaan-1 (Sridharan et al., 2010).

As discussed by Anand (2010) and Smith et al. (2012), obtaining improved knowledge of the presence, composition, and abundance of water (and other volatiles) at the lunar poles is important for several reasons:

- It is probable that the ice in permanently shadowed regions is ultimately derived from comet and/or asteroid impacts. Even though the original volatiles will have been considerably reworked, it is likely that some information concerning the composition of the original sources will remain. Among other things, this may yield astrobiologically important knowledge on the role of comets and meteorites in delivering volatiles and pre-biotic organic materials to the terrestrial planets (Chyba and Sagan 1992; Pierazzo and Chyba 1999).

- The processes involved in the creation, retention, migration, and destruction of OH and $H_2O$ across the surface of the Moon are likely to be common on other air-less bodies, and quantifying them on the Moon will give us better insight into the volatile history and potential availability of water elsewhere in the inner solar system.

- Lunar polar ice deposits are of considerable astrobiological interest, even if they do not retain vestigial information concerning their ultimate sources. This is because any such ices will have been subject to irradiation by galactic cosmic rays and, as such, may be expected to undergo organic synthesis reactions (e.g. Lucey, 2000). Analogous reactions may be important for producing organic molecules in the icy mantles of interstellar dust grains, and on the surfaces of outer Solar System satellites and comets (e.g. Bernstein et al., 2002; Elsila et al., 2007), but the lunar poles are much more accessible than any of these other locations.

- The presence of water ice at the lunar poles, and even hydrated materials at high-latitude but non-shadowed localities, could potentially provide a very valuable resource in the context of future human exploration of the Moon (e.g. Anand et al., 2012; this issue).

Confirming the interpretation of the remote sensing measurements, and obtaining accurate values for the concentration of polar ice and high latitude surficial OH/$H_2O$ will require *in situ* measurements by suitably instrumented and landed spacecraft. The requirements for permanently shadowed environments and high latitude, but non-permanently shadowed, environments are rather different. The conditions within permanently shadowed areas, where temperatures can be below 40 K (Paige et al., 2010), are not readily amenable to human exploration or even robotic vehicles (unless these are equipped with a nuclear power source or long-lived batteries capable of operating at very low temperatures). For these environments a penetrator-based system, such as envisaged in the MoonLITE and LunarNet mission concepts (Gao et al., 2008; Smith et al., 2012), may be a possibility. On the other hand, non-permanently shadowed polar localities *are* amenable to solar-powered robotic exploration, and proposals such as ESA's Lunar Lander (Carpenter et al., 2010; 2012), and 'Lunar Beagle' (Gibson et al., 2010), would provide valuable initial measurements of volatiles in these environments. In the longer term, a full characterization of polar volatiles, as for other aspects of lunar geology, would benefit from the increased mobility and flexibility that would be provided by human exploration (which would of course be facilitated at the poles if exploitable quantities of volatiles prove to be present).

**2.5 Lunar volcanism**

The SCEM Report identified the charaterization of lunar volcanism as a high lunar science priority because of the window it provides into the thermal and compositional evolution of the lunar mantle. Since the SCEM Report was published new remote sensing observations have indicated that lunar volcanism is even more diverse than previously thought, with the identification of probable areas of non-mare silicic volcanism on both the near- and far-sides (Glotch et al., 2010; Jolliff et al., 2011). There is no doubt that these areas would benefit from *in situ* investigation. However, from a lunar exploration perspective, this is just one aspect of the wider requirement to sample a diverse set of lunar rocks (Section 2.3), and the implications for exploration capabilities are essentially the same.

**2.6 Impact processes**

Impact cratering is a fundamental planetary process, an understanding of which is essential for our knowledge of planetary evolution. Yet our knowledge of impact processes is based on a combination of theoretical modelling, small-scale laboratory hyper-velocity impact experiments, and field geological studies of generally poorly-preserved terrestrial impact craters (Melosh, 1989). The Moon provides a unique record of essentially pristine impact craters of all sizes (from micron-sized pits up to

1000-km impact basins). Field studies, combining sample collection (including drill cores) and *in situ* geophysical studies (e.g. active seismic profiling), of the ejecta blankets and sub-floor structures of pristine lunar craters of a range of sizes would greatly aid in our understanding of the impact cratering process. As discussed by Crawford (2004), the implied requirements for mobility, deployment of complex geophysical instruments, sub-surface drilling, and sample return capacity are likely to outstrip the capabilities of purely robotic exploration and would be greatly facilitated by a human exploration programme.

**2.7 Regolith processes**

The SCEM Report noted that the lunar surface is a natural laboratory for understanding regolith processes and space weathering on airless bodies throughout the Solar System. In addition to their scientific interest, better characterisation of the composition, volatile content and mechanical properties of lunar regolith will also be important for planning In-Situ Resource Utilisation (ISRU) applications in support of future lunar exploration activities (e.g. Anand et al., 2012; Schwandt et al., 2012). As noted in Section 2.4, the nature of relatively cold high-latitude regoliths, which have never been sampled or studied *in situ*, and which may contain a volatile component, are of particular interest. Many of the properties of both high- and low-latitude regoliths could be investigated by appropriate instruments on robotic soft-landers such as ESA's proposed Lunar Lander (Carpenter et al., 2012) or penetrators such as the LunarNet concept (Smith et al., 2012).

Another important aspect of the lunar regolith is the record it contains of early solar and Solar System history. Studies of Apollo samples have revealed that solar wind particles are efficiently implanted in the lunar regolith (McKay et al., 1991; Lucey et al., 2006), which therefore contains a record of the composition and evolution of the Sun throughout Solar System history (e.g. Wieler et al., 1996). Recently, samples of the Earth's early atmosphere appear to have been retrieved from lunar regolith samples (Ozima et al., 2005; 2008), and it has been suggested that samples of Earth's early crust may also be preserved there in the form of terrestrial meteorites (Gutiérrez,, 2002; Armstrong et al., 2002; Crawford et al., 2008; Armstrong, 2010). Meteorites derived from elsewhere in the Solar System will likely also be found on the Moon, preserving a record of the dynamical evolution of small bodies throughout Solar System history (Joy et al., 2011; 2012). Last but not least, the lunar regolith may contain a record of galactic events, by preserving the signatures of ancient galactic cosmic ray (GCR) fluxes, and the possible accumulation of interstellar dust particles during passages of the Sun through dense interstellar clouds (Crozaz et al., 1977; McKay et al., 1991; Crawford et al., 2010). Collectively, these lunar geological records would provide a window into the early evolution of the Sun and Earth, and of the changing galactic environment of the Solar System, that is unlikely to be obtained in any other way. Much of this record has clear astrobiological implications, as it relates to the conditions under which life first arose and evolved on Earth.

From the point of view of accessing ancient Solar System history it will be desirable to find layers of ancient regoliths (*palaeoregoliths*) that were formed and buried billions of years ago, and thus protected from more recent geological processes, (e.g. Spudis, 1996; Crawford et al., 2010; Fagents et al., 2010; see Figure 1 of Crawford et al., 2010 for a pictorial representation of the process). Locating and sampling such deposits will likely be an important objective of future lunar exploration activities, but they will not be easy to access. Although robotic sampling missions might in principle be able to access palaeoregoliths at a limited number of favorable sites (for example where buried layers outcrop in the walls of craters or rilles), fully sampling this potentially rich geological archive of Solar System history will probably require the mobility and sample return capabilities of human exploration (Spudis, 2001; Garvin, 2004; Crawford, 2004). The specific requirements of a human exploration architecture capable of locating and sampling palaeoregolith deposits have been described elsewhere (Crawford et al., 2010).

**2.8 Atmospheric and dust environment**

The final broad lunar science area discussed by the SCEM Report related to studies of the lunar atmosphere and near-surface dust environment. Of particular interest is the need to characterise the composition of the tenuous lunar exosphere (which has a variable density in the range $10^5 - 10^7$ atoms or molecules cm$^{-3}$; Lucey et al., 2006), and its interaction with the surface regolith. This in turn will help constrain models of the lunar volatile budget discussed in Section 2.4. The extent to which transient releases of gasses into the atmosphere may occur is also of interest, as this may correlate with on-going low-level geological activity (Crotts, 2008). The surface dust environment, and especially the extent to which dust grains may become electrostatically charged and transported, is another important research topic (e.g. Grün et al., 2011; Pines et al., 2011; Horanyi and Stern, 2011), not least because of the potential hazards mobile dust may pose to scientific instruments and human operations on the surface (NRC, 2007; Loftus et al., 2010; Linnarsson et al., 2012). Some of the processes involved are likely to be common on other air-less bodies, and quantifying them on the relatively accessible lunar surface will therefore give us better insight into regolith/exosphere interactions throughout the Solar System.

Although some aspects of these investigations can be performed from lunar orbit (for example relevant observations will be performed by the Lunar Atmosphere and Dust Environment Explorer, LADEE, mission due for launch in 2013), more detailed studies of the lunar dust and plasma environment will require *in situ* surface measurements. These may be provided by suitably instrumented robotic landers, such as ESA's proposed Lunar Lander mission (Carpenter et al., 2012). However, it is important to note that the landed missions have the potential to significantly disturb the tenuous lunar atmospheric environment, and that this will be especially true of human operations (NRC, 2007). It is therefore scientifically highly desirable that the lunar atmosphere/exosphere be properly characterized by minimally invasive robotic probes before human operations are resumed on the lunar surface.

## 3. Science on the Moon

### 3.1 Life sciences and astrobiology

The Moon is a potentially valuable site to address a range of life science and astrobiology questions (e.g. Crawford, 2006; Gronstal et al., 2007; Cockell 2010; Crawford and Cockell, 2010). These questions broadly fall into three categories:

3.1.1 *Research that enhances our understanding of the habitability of the Earth through time*

As the Earth's closest celestial neighbour the Moon retains a unique record of the inner Solar System environment under which life evolved on our planet. The metamorphism and alteration of terrestrial Archaean (i.e. >2.5 Gyr old) rocks and their organic microfossils limits the quantity of material that can be used to understand the nature of early life on the Earth. The possibility that rocks ejected by asteroid and comet impacts on the early Earth may have landed on the Moon provides a tantalising possibility for a lunar surface source of early Earth material (Gutiérrez,, 2002; Armstrong et al. 2002; Crawford et al., 2008; Armstrong, 2010). The quantity of this material is predicted to be as much as 200 kg/km$^2$. The Moon may also have collected material ejected from other planetary bodies in the Solar System. Early rocks from Mars and Venus, both of which have early histories of enormous biological interest, might also exist on the Moon (Gladman et al. 1996; Armstrong et al. 2002), although the quantity and distribution and condition of this material is more uncertain. In addition, the lunar regolith, and especially buried palaeoregoliths (Spudis 1996), likely contains a record of solar wind flux (and thus solar luminosity) and galactic cosmic rays (and thus the galactic environment of the solar system) throughout solar system history (Crawford et al., 2010; Fagents et al., 2010). Much of this record will be directly relevant to understanding the past habitability of our own planet.

3.1.2 *Research that enhances our understanding of the possibility of life elsewhere in the Universe*

Although the Moon has, almost certainly, never supported any life of its own, lunar exploration will nevertheless inform our searches for life elsewhere. This includes a record of volatile fluxes in the inner solar system (NRC, 2007; Anand, 2010), and information on the survival of both microorganisms and organic matter in extreme planetary conditions.

The impacts that occurred into the Moon in its early history included material in addition to terrestrial rocks, such as cometary material and chondritic and carbonaceous meteorites (e.g. Joy et al., 2012). Protected either in the subsurface or in permanently shadowed craters (Seife, 2004; Vasavada et al., 1999), these might provide insights into the inventory of volatile and/or organic material that penetrated the early inner Solar System, and what quantity may still do so today. Insofar as these organics might have provided an exogenous source of prebiotic organics necessary to kick-start life on the Earth, investigating organics on the Moon has important contributions to make to understanding the origin of life on the Earth. Indigenous organic processing on the Moon may also yield insights into the chemical pathways of alteration, and fate, of organics in interplanetary space (e.g., Lucey, 2000).

In addition, the lunar environment contains the crashed remains of unsterilized spacecraft. The bacterial spores that these craft may contain could be collected and examined for DNA and other biochemical damage, as well as examined for their viability. Human or robotic explorers could be developed to collect these organisms for study on the Moon or return to the Earth. The organisms returned from these craft would answer many questions about the longevity of microorganisms in the space environment that will inform fields as diverse as planetary protection, for example allowing for an assessment of how long contaminant organisms survive on other planetary surfaces (Rummel, 2004; Glavin et al., 2004, 2010) and biogeography, for example, showing whether organisms can survive the conditions in interplanetary space and the impact conditions of landing on another planetary surface after being transferred from one planet to another (Clark, 2001; Horneck et al., 2001; Mastrapa et al., 2001; Burchell, 2004; Cockell et al., 2007; Nicholson et al., 2005).

3.1.3 *Research that advances the human exploration and settlement of space*

The space environment is hostile to life and includes hard vacuum, high radiation (both UV and ionizing radiation), altered gravity regimes, the presence of biologically and mechanically aggravating dust, and difficulties in acquiring liquid water and gases to breathe (e.g. Horneck, 1996; Horneck et al., 2003). The Moon is therefore a testing ground for technological principles and approaches for dealing with the major environmental parameters that affect life in outer space. For example, the Moon can be used to investigate whether the effects of gravity are linear or whether there are critical threshold in effect (e.g. Cockell, 2010), the biological effects of the radiation environment beyond the Earth's magnetosphere, and the toxicity of lunar dust (Carpenter et al., 2010; Loftus et al., 2010; Linnarsson et al., 2012).

Organisms could also be taken to the lunar surface and used to carry out investigations *in situ* using surface laboratories (as proposed for ESA's Lunar Lander; Carpenter et al., 2010). Microorganisms, plants and animals could be used to investigate a variety of questions, including the cumulative effects of space conditions (Mileikowsky et al., 2000; Blakely, 2000; Clark, 2001; Brenner et al., 2003; Giusti et al., 1998; Horneck et al., 2003; Stein and Leskiw ,2000; Zayzafoon et al., 2005; Carpenter et al., 2010) and their use in life support systems (Tamponnet 1996; Sadeh and Sadeh, 1997; Bluem and Paris, 2001; Henrickx et al., 2006). These experiments would yield new insights into the evolution of organisms in the space environment, the possibility of microbial, plant, and cultivated crop production, and the potential for healthy human and animal reproduction in space.

**3.2 Human physiology and medicine**

Acceleration produced by the force of gravity is an omnipresent factor that modulates many biological processes (Clément and Slenzka, 2006). Historically, much of the space biomedical science focus has been upon whole-body physiology, reflecting the immediate challenge of maintaining astronaut (and thus mission) functionality (Garshnek, 1989; Williams, 2003). Most physiological systems are known to progressively 'adapt' to life in microgravity (Nicogossian et al., 1994), however many encounter problems upon re-exposure to a gravitation vector. Alterations in spatial orientation (Lipshits et al., 2005) and sensory-motor function (e.g. Kalb and Solomon, 2007; Souvestre et al., 2008) are acutely challenging but re-adapt rapidly, whereas cardiovascular (Hargens and Richardson, 2009; Hughson, 2009) and musculoskeletal system (Narici and de Boer 2011) de-conditioning can precipitate chronic health issues as well as imparing astronaut operations. A raft of countermeasures including various exercise regimes, nutrition and behavioural support are employed in space missions (Convertino, 2002; Cavanagh et al., 2005), but these fail to entirely ameliorate microgravity-induced de-conditioning. As a consequence, the maximum 'recommended' stay on the International Space Station (ISS) is currently 6 months (Williams et al., 2009).

Partly, this failure relates to the fact that just *how* gravity affects biological processes is largely unknown, at the level of an organism, a system, or of an individual cell. In fact, whilst life appears well adapted to gravity today, it presented a major challenge to emergence from the prehistoric aquatic environment. Gravitropism (a plant growth's sensitivity to gravity) has been well documented since Darwin, and it is highly likely that gravity also has an important role in the regulation of animal cells. In fact, intra-cellular force sensors (sensitive to physical pressure and structural strain) have been proposed as part of standard cellular architecture (Wayne et al, 1992).

Even minor alterations in the physical force environment (Klaus, 1998) may have significant downstream effects. An abnormal gravity environment is likely to inhibit sedimentation, equalising molecular (and thus affecting electro-chemical gradients) and organelle distribution, and modify cytoskeletal activity and gene transcription (Cogoli and Cogoli-Greuter, 1997). Such processes may in part explain observed gravity-dependent changes in cellular growth, proliferation and regeneration (Sonnenfeld and Shearer, 2002; Borchers, et al., 2002), organisation (Vico et al., 2000), and healing (Davidson et al., 1999; Radek et al., 2008). For instance, reduced lymphocyte proliferation, delayed bone cell differentiation (Hughes-Fulford and Lewis, 1996), and retardation of pre-natal (e.g. Bruce, 2003) and post-natal development (e.g. Ronca and Alberts, 1997) and locomotion (Walton, 1998) have all been observed in a number of animal models in microgravity.

Monitoring human adaptation to prolonged exposure to partial gravity, such as exists on the Moon, may offer significant insights into vestibular disorders (Clément et al., 2005) and a range of processes beyond associated in ageing (Vernikos and Schneider, 2010), disuse pathology (Edgerton et al., 2000; Elmann-Larsen and Schmitt, 2003) and lifestyle conditions such as the metabolic syndrome and cardiovascular disease. The lunar partial gravity model would supplement the knowledge accrued by 'long duration' ISS microgravity exposure, by adding a point at $1/6^{th}$ g from which to investigate gravity dose-dependence (e.g. Cockell, 2010). In addition, by providing a more tightly controlled environment in terms of the nutritional, medical (including remote monitoring) and exercise countermeasure support, the Moon would offer an opportunity to test the efficacy of countermeasure and/or therapeutic approaches within otherwise 'healthy' and well motivated individuals (e.g. Green, 2010). The lunar partial gravity is likely to offer some protection to de-conditioning that will likely not only facilitate longer operational durations than the ISS, but also has greater relevance and applicability to life on Earth than the microgravity environment of the ISS. There would therefore be the opportunity to understand the physiological effects of gravity that are not possible within either a 1-g or a microgravity environment. Such studies were beyond the scope of the limited duration Apollo missions, but would be facilitated by longer-term human operations on the Moon.

Finally, there would be much to learn about life support (e.g. bio-regenerative food, breathable air, and water closed-loops; Ekhart, 1996), and medical support provision, from human operations in a lunar base beyond research into partial gravity effects. Examples include individualised medicine (Kalow 2002), carcinogenesis (Rykova et al., 2008), viral virulence (Wilson et al., 2007), robotic surgery (Cermack, 2006), telemetric medical monitoring (Grigoriev and Egorov, 1997), and even basic health care delivery due to the severe resource and technical assistance limitations (Mortimer et al., 2004). In contrast to the popular myth, space biomedical devices must be simple (to operate and repair), robust, compact, light, low consumers, non-invasive and multi-faceted (e.g. diagnostic and interventional ultrasound; Ma et al., 2007) and ideally preventative (Thirsk et al., 2009). Such characteristics possess value well beyond space applications, e.g. in remote societies and disaster zones, and could also facilitate de-centralised care provision within the developed world offering efficacy and cost-effectiveness benefits (Williams, 2002).

It follows that while incorporating humans in future lunar exploration may add to the associated risks and challenges, it would radically enhance both its scientific capabilities and the resulting benefits (including potential biomedical benefits) to society.

**3.3 Fundamental physics**

Although not a major driver for lunar exploration, it is recognized that a number of research fields in the area of fundamental physics may benefit from the ability to place scientific instruments on the lunar surface. These include tests of General Relativity through improved lunar laser ranging measurements (e.g. Livio, 2006; Burns et al., 2009; Currie et al., 2010), tests of quantum entanglement over large baselines (Schneider, 2010), and searches for strange quark matter (Banerdt, et al., 2007; Han et al., 2009).

**4. Science from the Moon – Astronomy**

**4.1 Why Astronomy?**

A natural area to use the Moon as a platform for performing scientific experiments is astronomy (for summaries see, e.g., Burns et al., 1990; Livio, 2006; Crawford and Zarnecki, 2008; Jester and Falcke, 2009). Almost the entire electromagnetic spectrum is currently being used to study the universe from radio to high-energy gamma ray emission. Different frequencies typically relate to different physical processes, and consequently the universe looks markedly different in optical, infrared, or radio wavelengths. Hence, during the last century modern telescopes have diversified and evolved enormously, fundamentally changing our view of the universe and our place therein. Due to their ever increasing sensitivity, which allows one to peer deeper and deeper into the earliest phases of the cosmos, the requirements for telescope sites have become more and more extreme: one simply needs the best possible observing conditions. The most important factors here are light pollution (at the relevant frequencies) and distortions due to the atmosphere. Light pollution is generally caused by any form of civilization, thereby pushing observatories to more and more remote locations. Detrimental effects of the atmosphere include:

- temporary effects such as clouds and water vapour, which temporarily absorb and disturb optical or high-frequency radio radiation,
- turbulence in the ionosphere or troposphere, which distorts radio or optical wave fronts, thereby severely degrading the image quality,

- air glow, which can overpower sensitive infrared observations,
- total absorption of radiation, e.g., of very low-frequency radio, infrared, X-ray, and gamma-ray radiation.

The best – and in many cases only – remedy is to observe from dry deserts, high mountains, or from space. Two of the most remote, but also most exquisite, astronomical sites on Earth are the Atacama desert and Antarctica. The former currently hosts some of the world's largest telescopes, including ESO's 8m-class Very Large Telescopes (VLT), the ALMA sub-mm-wave radio telescope, and in the future probably also the ~40 m diameter European Extremely Large Telescope (E-ELT; see http:// www.eso.org). A century after its initial exploration, Antarctica now also hosts a number of somewhat smaller telescopes (e.g., the South Pole Telescope, Carlstrom et al., 2011) as well as the giant IceCube detector. IceCube is the world's largest neutrino observatory, using the ice itself as detector material (e.g., Abbasi et al., 2011).

The Moon would be a logical next step in the quest for the most suitable sites to be used for astronomy. An important secondary important factor in selecting a site, however, is the available infrastructure: How accessible is the site for people and material? How does one obtain power and how good is the data connection? Already for Antarctica this poses serious constraints, and it took a long time until this continent became useful for scientific exploitation. It is needless to say that the Moon is even more difficult to reach. Hence, like Antarctica, any significant exploitation of the Moon requires a developed infrastructure – something that would likely become available only in conjunction with human exploration of the Moon. Even then one has to assess how unique and useful the Moon is for astronomy in the first place. After all, the International Space Station (ISS), while having a well-developed infrastructure available, is not used for telescopes; its small, relatively unstable platform in low-Earth orbit (LEO) is simply too poor a telescope site to be competitive. Hence, the vast majority of space-based telescopes have been associated with free-flying satellites. Of course, some of these satellites, most notably the Hubble Space Telescope (HST), benefited from the heavy lift capabilities of the Space Shuttle and the servicing possibilities the human space flight program offered (NRC, 2005). Indeed, it is interesting to note that the one human-serviced space telescope, HST, is in fact the most productive of all astronomy space missions even many years after its launch (see Tables 4 and 6 in Trimble and Ceja, 2008; HST produced 1063 papers in the time frame 2001-2003, compared to 724 for Chandra, the next most productive).

So, the question to ask is: Which type of telescopes would uniquely benefit from a lunar surface location? This question has been addressed in a couple of workshops and scientific roadmaps in recent years (Falcke et al., 2006; Livio, 2006; NRC, 2007; Crawford and Zarnecki, 2008; Worms et al., 2009). In the following section we try to synthesize these findings.

**4.2 Which astronomy?**

There is a wide consensus that a low-frequency radio telescope (i.e. a radio telescope operating at frequencies below 30-100 MHz) would be the highest priority (e.g., Jester and Falcke, 2009; Burns et al., 2009). Radio waves at these frequencies are seriously distorted by the Earth's ionosphere and completely absorbed or reflected at frequencies below 10-30 MHz. Hence, the low-frequency universe is the last uncharted part of the electromagnetic spectrum, and a lunar infrastructure would greatly benefit its exploration. Of particular relevance for science here is the investigation of the "dark ages" of the universe. This is the epoch several hundred million years after the big bang, but before the formation of the first stars and black holes, when the cosmos was mainly filled with dark matter and neutral hydrogen. This epoch contains still pristine information of the state of the big bang and can essentially only be observed through radio emission from atomic hydrogen red-shifted to several tens of MHz. The best location to study this treasure trove of cosmology (Loeb and Zaldariaga 2004) would indeed be on the lunar far-side.

Other science topics of interest for a lunar radio telescope include:

- A high-resolution map of the universe and a general inventory of low-frequency radio sources – ground-based maps around 10 MHz are very poor compared to any other wavelength due to the effects of the ionosphere (Cane and Whitham 1977);

- Search for and study of radio emission from planets and exoplanets (Grießmeier et al. 2007);

- Investigation of the local plasma bubble around our solar system; and

- Radio-detection of ultra-high energy cosmic rays. In fact the Moon itself is already now being used as a detector for ultra-high energy neutrinos (Gorham et al. 2004, Buitink et al. 2010, Jaeger et al. 2011) and cosmic rays (ter Veen et al. 2010) with the help of radio telescopes.

More details on the various science cases, an extensive review of past studies, and a summary of relevant observing constraints can be found in Jester and Falcke (2009).

Low-frequency telescopes are currently being built using a network of many, relatively simple dipole-like antennas (e.g., the LOFAR telescope; van Haarlem et al. 2012). The individual antennas are digitally connected to form a large interferometer (a "phased array"), which acts as one large, steerable telescope dish, but without any moving parts. These radio antennas can be made robust and lightweight to transport many of them into space, while the Moon's gravity and surface would keep them fixed relative to each other. Moreover, if placed at the far-side of the Moon, the Moon itself would shield the telescope from any radio pollution (also called radio frequency interference, RFI) originating from Earth. This RFI can be either of man-made origin or due to natural Auroral Kilometric Radiation (AKR) in the Earth's magnetosphere. In fact – due to this shielding and lack of its own magnetic activity – the farside of the Moon probably belongs to the most radio-quiet places in our solar system. This has been long recognized and the International Telecommunication Union (ITU) has accordingly designated the lunar far-side as a radio-quiet zone for radio astronomy (ITU Radio Regulations, Article 22, Section V). The installation of such a low-frequency radio array could proceed in several steps, starting with a few antennas and then growing to a larger and larger network (ideally out to tens to hundreds or kilometres with thousands of antennas). While the first steps can certainly be done robotically, a large-scale installation over such diverse terrain and large distances, will ultimately be much more efficient with the presence of humans on the surface.

In addition to radio telescopes, a number of other types of lunar observatories have been discussed (for example, see the discussion reported by Crawford and Zarnecki, 2008). The huge cost overruns of the HST-successor, JWST, have shown how difficult it is to fold-up and bring even medium-sized optical telescopes into space. For certain applications liquid-mirror telescopes have been proposed, whereby the combination of a rotating fluid and the Moon's gravity would form a perfect (non-steerable) mirror that could detect the very first stars in the universe (Angel et al. 2008), although this technique currently has a rather low technical readiness. Perhaps a simpler, and scientifically topical and important, implementation of lunar-based telescopes would be to obtain disc-integrated spectral, polarimetric, and albedo measurements of the Earth. Such observations could provide important insights into both the Earths radiation budget and climate (e.g. Pallé and Goode, 2009), and the interpretation of observations of Earth-like exoplanets and the interpretation of planetary biosignatures (e.g. Sparks et al., 2010; Karalidi et al., 2012).

With regard to larger instruments, the lunar surface and gravity may facilitate the construction of interferometers by combining a set of smaller optical, IR, and sub-mm-wave telescopes – a concept similar to the radio interferometer array. However, here shielding is not such an issue and one needs to compare these Moon-based interferometers with free-flying and formation-flying satellites. The latter concept has been the preferred option for small-number-of-elements interferometers (e.g., Cockell et al. 2009). On the other hand, formation flight to the required precision for astronomical interferometers has not been realized so far, while ground-based interferometers are well understood and tested.

Moreover, the Moon as a large airless body could host detectors for cosmic ray particles. Those particles are typically absorbed in the Earth's atmosphere but would reach the lunar surface unimpeded. The detectors could be distributed over the Moon to study solar-wind induced structural variations of the cosmic ray flux at low energies. At energies above $10^{15}$ eV the composition of cosmic rays could be directly measured, if a large detector can be built using, e.g., lunar water resources. Here, however, one needs to critically assess whether the science return justifies the undoubtedly high costs for these concepts.

Finally, the simple presence of a lunar infrastructure for human activities, may lend itself to the relatively simple installation of smaller telescopes and experiments. Indeed, Apollo 16 deployed a UV telescope on the Moon (Carruthers, 1973) – a class of telescopes that was later flown on satellites (e.g. the International Ultraviolet Explorer, IUE). This would qualify as "opportunistic" science, i.e., missions that simply catch a ride, given it is there. At the very least, an existing 'science-park' on the Moon would take the burden of taking along one's own attitude control, power-supply, or data handling capability, which now has to be carried by every astronomy satellite. Whether by the time such an infrastructure is likely to be available there will still a sufficient science case left for such small telescopes remains to be seen. On the other hand, it is worth noting that, perhaps surprisingly, small robotic telescopes have revolutionized some areas of astronomy (e.g., planet searches, Gamma-Ray-Burst afterglows) in recent years and may continue to do so.

To summarize: a low-frequency lunar radio telescope to map the last unexplored frequency window to the universe, and to explore the dark ages of the universe, should be feasible and would be uniquely suited for the Moon's far-side. Small optical/IR telescopes to observe the Earth from the nearside may also be scientifically valuable. Large (10 metre plus) optical telescopes or interferometers might at some point benefit from the lunar surface as a platform, but the technical case for this still needs to be developed, and the Moon as a site for such telescopes needs to be better studied. Smaller telescopes could be installed, using lunar activities as opportunity. All options would certainly benefit from a sustainable human return to the Moon, but certainly will not drive it. This situation is, however, not very much different from the usage of Antarctica for astronomy, where astrophysicists simply made use of the fact that Antarctic exploration programs and infrastructure already existed. Perhaps the same will happen with the Moon in the coming decades.

As a final note: a potentially disturbing factor for many telescopes at wavelengths shorter than radio is the unknown dust mobility on the Moon, which might cover mirrors and joints over a period of time. Also, for radio telescopes the dielectric properties of the lunar surface, the existence of an ionosphere, and the temporal (e.g., through cosmic rays) and spectral radio background variations should be studied to verify the feasibility of the lunar surface for future astronomy applications. Hence, any lunar precursor missions, such as ESA's proposed Lunar Lander (Carpenter et al., 2008; 2012) should investigate these effects in more detail in order to optimize future astronomical observations from the Moon..

## 5 Conclusions

Summarising the above, we see that the lunar geological record still has much to tell us about the earliest history of the Solar System, the origin and evolution of the Earth-Moon system, the geological evolution of rocky planets, and the near-Earth cosmic environment throughout Solar System history. These lunar science objectives were strongly endorsed by the US National Research Council Report on the Scientific Context for Exploration of the Moon (NRC 2007), and a number of them (i.e. those which could be addressed by sample return from the South Pole-Aitken Basin and by the creation of a lunar geophysics network) received high priority in the recent US Planetary Science Decadal Survey (NRC, 2011). Addressing these objectives requires an end to the 40-year hiatus of lunar surface exploration, with the placing of scientific instruments on, and the return of samples from, the surface of the Moon, with a particular emphasis on regions not previously visited. It is also clear that the lunar surface offers outstanding opportunities for research in astronomy, astrobiology, fundamental physics, life sciences and human physiology and medicine.

Many of these objectives, can be addressed robotically, as reflected by the large number of proposals for lunar surface robotic exploration that have been put forward in recent years (e.g. MoonLITE and MoonRaker, Gao et al., 2008; MoonRise, Jolliff et al., 2010; Lunar Beagle, Gibson et al., 2010; SELENE-2, Hashimoto et al., 2011; Luna-Glob, Mitrofanov et al., 2011; Lunette, Neal et al., 2011; LunarNet, Smith et al., 2012; Farside Explorer, Mimoun et al., 2012; and ESA's Lunar Lander, Carpenter et al., 2012). However, in the longer term, it is also clear that most of these scientific objectives would benefit from the scientific infrastructure, on the spot decision making, enhanced surface mobility, and sample return capacity that would be provided by renewed human operations on the lunar surface (e.g. Spudis, 2001; Garvin, 2004; Cockell, 2004; Crawford, 2004; 2012). Indeed, some of these scientific objectives will be impossible to conduct robotically (i.e. those related to human physiology and medicine where humans will form the experimental subjects), and others (e.g. deep drilling into the lunar crust to extract undisturbed palaeoregolith deposits with their potentially rich record of Solar System history) may be wholly impractical without a human presence. That said, it is also true that a human return to the Moon will benefit from robotic precursor missions, for example to assess possible seismic and impact hazards, regolith properties (including possible dust toxicity; Linnarsson et al. 2012, this issue), the radiation environment, and *in situ* resource availability (Anand et al. 2012, this issue). For all these reasons it is highly desirable that current plans for robotic exploration of the lunar surface are developed in the context of a future human exploration programme.

Fortunately, such a programme is under active international discussion. In 2007 the World's space agencies came together to develop the Global Exploration Strategy (GES), which lays the foundations for a global human and robotic space exploration programme (GES 2007). One of the first fruits of the GES has been the development of a Global Exploration Roadmap (GER 2011), which outlines possible international contributions to human and robotic missions to the Moon, near-Earth asteroids and, eventually, Mars. Implementation of this roadmap would provide many opportunities for pursuing the science objectives outlined in this paper, and lunar science would therefore be a major beneficiary of its implementation.

**Acknowledgements**

We wish to thank Dr James Carpenter for his enthusiastic support of lunar science in Europe, for organising this Special Issue of Planetary and Space Science, and for his invitation to contribute to it. We thank the two referees (Dr. Wim van Westrenen, the other anonymous), for comments which have greatly improved the quality of the manuscript.

Bruce, L.L., 2003. Adaptations of the vestibular system to short and long-term exposures to altered gravity. *Adv Space Res*. 32(8), 1533-9.

Bruce, L.L., 2003. Adaptations of the vestibular system to short and long-term exposures to altered gravity. *Adv Space Res*., 32(8), 1533-9.

Buitink, S., Scholten, O., Bacelar, J., Braun, R., de Bruyn, A.~G., Falcke, H., Singh, K., Stappers, B., Strom, R.G., Yahyaoui, R.A., 2010. Constraints on the flux of ultra-high energy neutrinos from Westerbork Synthesis Radio Telescope observations. *Astron. Astrophys*., 521, A47.

Burchell, M.J., 2004. Panspermia today. *Int. J. Astrobiology* 3, 73–80.

Burns, J.O., Duric, N., Taylor, G.J., Johnson, S.W., 1990. Observatories on the Moon. *Scientific American*, 262(3), 18-25.

Burns, J.O., et al., 2009. Science from the Moon: The NASA/NLSI Lunar University Network for Astrophysics Research (LUNAR). Submitted as a white paper to the planetary sciences decadal review, ArXiv e-prints arXiv:0909.1509.

Cane, H.V., Whitham, P.S., 1977. Observations of the southern sky at five frequencies in the range 2-20 MHz. *Mon. Not. Roy. Astron. Soc*., 179, 21-29.

Carlstrom, J.E., et al., 2011. The 10 Meter South Pole Telescope. *Pub. Astron. Soc. Pacific*, 123, 568-581.